\documentclass[twocolumn,english,showpacs,showkeys]{revtex4}
\usepackage[T1]{fontenc}
\usepackage[latin9]{inputenc}
\usepackage{color}
\usepackage{float}
\usepackage{textcomp}
\usepackage{graphicx}
\usepackage{amssymb}
\usepackage{esint}

\DeclareRobustCommand{\greektext}{%
  \fontencoding{LGR}\selectfont\def\encodingdefault{LGR}}
\DeclareRobustCommand{\textgreek}[1]{\leavevmode{\greektext #1}}
\DeclareFontEncoding{LGR}{}{}

\usepackage{babel}

\begin{document}

\title{\[
\]
Thermal and Transport Behavior of Single Crystalline R$_{2}$CoGa$_{8}$
\\
(R = Gd, Tb, Dy, Ho, Er, Tm, Lu and Y) Compounds}

\author{Devang A. Joshi, A. K. Nigam, S. K. Dhar and A. Thamizhavel}

\affiliation{Department of Condensed Matter Physics and Material Sciences, Tata
Institute of Fundamental Research, Homi Bhabha Road, Colaba, Mumbai
400 005, India.}
\begin{abstract}
The anisotropy in electrical transport and thermal behavior of single
crystalline R$_{2}$CoGa$_{8}$ series of compounds is presented.
These compounds crystallize in a tetragonal structure with space gropup
P4/mmm. The nonmagnetic counterparts of the series namely Y$_{2}$CoGa$_{8}$
and Lu$_{2}$CoGa$_{8}$show a behavior consistent with the low density
of states at the fermi level. In Y$_{2}$CoGa$_{8}$, a possibility
of charge density wave transition is observed at $\approx$ 30 K.
Gd$_{2}$CoGa$_{8}$ and Er$_{2}$CoGa$_{8}$ show a presence of short
range correlation above the magnetic ordering temperature of the compound.
In case of Gd$_{2}$CoGa$_{8}$, the magnetoresistance exhibits a
significant anisotropy for current parallel to {[}100{]} and {[}001{]}
directions. Compounds with other magnetic rare earths (R = Tb, Dy,
Ho and Tm) show the normal expected magnetic behavior whereas Dy$_{2}$CoGa$_{8}$
exhibits the possibility of charge density wave (CDW) transition at
approximately same temperature as that of Y$_{2}$CoGa$_{8}$. The
thermal property of these compounds is analysed on the basis of crystalline
electric field (CEF) calculations. 
\end{abstract}



\maketitle

\section{INTRODUCTION}

\textcolor{black}{Recently we reported the magnetic properties of
single crystalline R$_{2}$CoGa$_{8}$ series of compounds, inferred
from the thermal and field dependence of magnetization \citet{Devang}.
These compounds form only with the heavy rare earths (R = Gd to Lu),
in contrast to the iso-structural indides R$_{2}$CoIn$_{8}$\citet{Devang1}\ 
where the phase forms for all the rare earths except for La, Yb and
Lu. Y$_{2}$CoGa$_{8}$ and Lu$_{2}$CoGa$_{8}$ show diamagnetic
behavior pointing out a relatively low density of states at the Fermi
level. R$_{2}$CoGa$_{8}$ with magnetic rare earths order antiferromagnetically
at low temperatures with the highest N$\grave{\mathrm{e}}$el temperature
T$_{N}$ = 28 K in Tb$_{2}$CoGa$_{8}$. The magnetic ordering temperatures
are less compared to their corresponding indides. The easy axis of
magnetization for Tb$_{2}$CoGa$_{8}$, Dy$_{2}$CoGa$_{8}$ and Ho$_{2}$CoGa$_{8}$
was found to be along the {[}001{]} direction whereas for Er$_{2}$CoGa$_{8}$
and Tm$_{2}$CoGa$_{8}$ the easy axis changes to the basal plane
namely (100). Gd$_{2}$CoGa$_{8}$ having the }\textit{\textcolor{black}{S}}\textcolor{black}{-state
ion Gd$^{3+}$ was found to show isotropic magnetic behavior. A point
charge model calculation of the crystal electric field (CEF) effects
gave a qualitative explanation of the magnetocrystalline anisotropy
in this series of compounds and the appreciable deviation of the ordering
temperatures in Tb$_{2}$CoGa$_{8}$ and Dy$_{2}$CoGa$_{8}$ from
that expected on the basis of de-Gennes scaling. The aim of the present
paper is to investigate in detail the heat capacity and electrical
transport properties of R$_{2}$CoGa$_{8}$ compounds} \textcolor{black}{to
get more information about the crystal electric field effects, Schottky
contribution to the heat capacity, entropy associated with the magnetic
ordering, etc. The magnetoresistivity was also studied keeping in
mind the anomalously high magnetoresistance ($\sim$2700 \% at 2 K)
of Tb$_{2}$CoIn$_{8}$\citet{Devang1}.}

\section{EXPERIMENTAL}

Single crystals of R$_{2}$CoGa$_{8}$ compounds were grown using
Ga flux \textcolor{black}{as \ desc}ribed elsewhere\citet{Devang}.
An energy dispersive X-ray analysis (EDAX) was performed on all the
obtained single crystals to estimate the actual crystal composition.
The EDAX results confirmed the crystals to be of the stoichiometric
composition 2:1:8. To check for the phase purity, powder x-ray diffraction
pattern of all the compounds were recorded by powdering a few small
pieces of single crystal followed by the Rietveld analysis of the
obtained pattern. For the anisotropic transport measurements, the
single crystals were oriented along the principle directions viz.,
{[}100{]} and {[}001{]} by Laue back reflection method. The crystals
were cut to the required size for resistivity and heat capacity measurements
using a spark erosion wire cutting machine. The heat capacity, resistivity
and magnetoresistance measurements were performed using physical property
measurement system (PPMS - Quantum Design). \textcolor{black}{The
AC susceptibility of Gd$_{2}$CoGa$_{8}$ and Tb$_{2}$CoGa$_{8}$
was also measured in MPMS - Qunatum Design.}

\section{Results and Discussion}

\subsection{Y$_{2}$CoGa$_{8}$ and \textcolor{black}{Lu$_{2}$CoGa$_{8}$}}

\textcolor{black}{We first present the data on Y$_{2}$CoGa$_{8}$
and Lu$_{2}$CoGa$_{8}$, \ which are the nonmagnetic analogs of
the R$_{2}$CoGa$_{8}$ \ compounds, Co being non-magnetic in this
family of compounds.} \textcolor{black}{As mentioned above Y$_{2}$CoGa$_{8}$
and Lu$_{2}$CoGa$_{8}$ show diamagnetic behavior, in contrast to
the indide Y$_{2}$CoIn$_{8}$\citet{Devang1}, which is Pauli-paramagnetic.
The diamagnetic contribution arises from the filled electronic shells,
which dominate a modest Pauli-paramagnetic contribution arising from
a low density of the conduction electron states at the Fermi level
\ $N(E_{F})$. The evidence for low $N(E_{F})$\  comes from the
low temperature (1.8 to 10 K) heat capacity data which furnish $\gamma$
values of 2 and 4 mJ/mole K$^{2}$ for Y$_{2}$CoGa$_{8}$ and Lu$_{2}$CoGa$_{8}$
\ respectively (Fig. 1a inset). These \ values are low in comparison
to the corresponding indide Y$_{2}$CoIn$_{8}$ (12 mJ/mole-K$^{2}$).
An estimate of the density of states at the Fermi level is obtained
using the free electron relation}

\textcolor{black}{\begin{equation}
\gamma=\frac{2}{3}\:\pi^{2}\: k_{B}^{2}\: N(E_{F})\end{equation}
where $k_{B}$ is the Boltzmann constant. Substituting the value of
$\gamma$ = 2 mJ/mole-K$^{2}$, the density of states in Y$_{2}$CoGa$_{8}$,
for example, is found to be 1.6 x 10$^{35}$ erg$^{-1}$mole$^{-1}$
or 5.8 Ry$^{-1}$atom$^{-1}$.} %
\begin{figure}[h]
\includegraphics[width=0.4\textwidth]{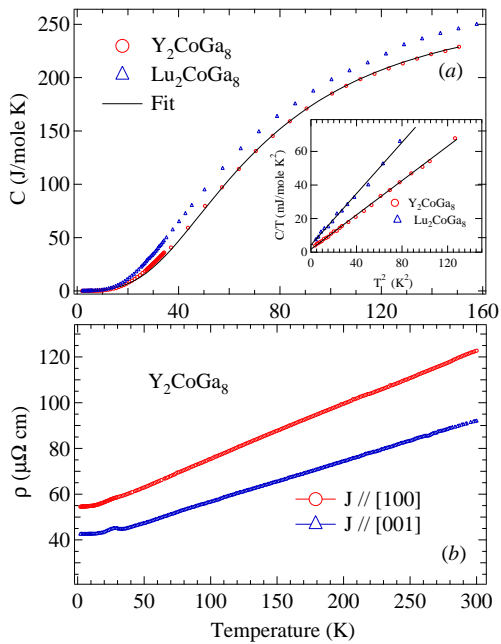}\caption{(a) 
Heat capacity curve of Lu$_{2}$CoGa$_{8}$ and Y$_{2}$CoGa$_{8}$
with a fit described in text. The inset shows the C/T \textit{vs}
T$^{2}$ plot. (b) Resistivity of Y$_{2}$CoGa$_{8}$ with the current
parallel to {[}100{]} and {[}001{]} directions.}

\end{figure}
\textcolor{black}{This value, for example, is comparable to that obtained
in the diamagnetic compound YPd$_{3}$\citet{Koenig} from band structure
calculations. In the free electron approximation $\gamma$ \ and
the Pauli susceptibility $\chi_{0}$ are related by the relation:
$\gamma(mJ/mole\, K^{2})=1.3715\times10^{-5}\chi_{0}(emu/mole)$.
For Y$_{2}$CoGa$_{8}$ and Lu$_{2}$CoGa$_{8}$ \ the gamma values
furnish $\chi_{0}$ less than $10^{-4}$ emu/mol. On the other hand
using the relation for diamagnetic contribution $\chi_{Dia}$ $\approx$
$-10{}^{-6}$ Z, where Z is the atomic number, we infer $\chi_{Dia}$
values of $-3.5\times10^{-4}$ and $-4.2\times10^{-4}$ emu/mole for
Y and Lu compounds, respectively, which in absolute magnitude are
higher than $\chi_{0}$. Replacing indium fully by gallium results
in the appearence of diamagnetism. Since the valency of both In and
Ga is same (both of them have one extra }\textit{\textcolor{black}{p}}\textcolor{black}{
\ electron), it is possible that shrinking of the unit cell in gallides
results in shifting of the Fermi level to a region of low density
of states. }

\textcolor{black}{From the slope of $C/T\,\, vs\,\, T^{2}$ plots
(inset Fig. 1a) we infer the lattice heat capacity coefficient $\beta$
as 0.501 and 0.782 $mJ/mole\, K^{4}$ in Y$_{2}$CoGa$_{8}$ and Lu$_{2}$CoGa$_{8}$,
respectively. Using the relation of the Debye model $\Theta_{D}^{3}=1943600/\beta$,
where $\beta$ is in the units of $mJ/g\, atom\, K^{4}$ and $\Theta_{D}$
is the Debye temperature, we obtain $\Theta_{D}$ = 349 and 301 K
in Y$_{2}$CoGa$_{8}$ and Lu$_{2}$CoGa$_{8}$, respectively.}\textcolor{blue}{
}\textcolor{red}{\ }\textcolor{black}{Based on the Debye approximation\citet{Hofmann},}

\textcolor{black}{\begin{equation}
\frac{\Theta_{D}(Y_{2}CoGa_{8})}{\Theta_{D}(Lu_{2}CoGa_{8})}=\sqrt{\frac{MW(Lu_{2}CoGa_{8})}{MW(Y_{2}CoGa_{8})}}\end{equation}
the ratio of the Debye temperatures is 1.16 in fair agreement with
the r.h.s value of 1.4. }

\textcolor{black}{The main panel of Fig. 1 shows the heat capacity
of Lu$_{2}$CoGa$_{8}$and Y$_{2}$CoGa$_{8}$ from 1.8 to 160 K.
The heat capacity curve for Y$_{2}$CoGa$_{8}$was fitted to the equation}

\textcolor{black}{\begin{equation}
C_{Tot}=\gamma T+C_{Ph}\end{equation}
where the two terms represents the electronic and phononic contributions,
respectively. C$_{Ph}$ can be written in terms of Debye integral
as }

\textcolor{black}{\begin{equation}
C_{Ph}=9NR\left(\frac{T}{\Theta_{D}}\right)^{3}\intop_{0}^{\Theta_{D}/T}\frac{x^{4}e^{x}dx}{\left(e^{x}-1\right)^{2}}\end{equation}
}

where $x$= $\Theta_{D}/T$, $N$ is the number of atoms in the formula
unit and $\Theta_{D}$ is the Debye temperature.\textcolor{blue}{
}\textcolor{black}{\ Here $\gamma$ was fixed to the values derived
above and $N$ = 11. The best fit shown by solid line in Fig. 1 is
obtained with $\Theta_{D}$ = 291 K. This value of $\Theta_{D}$ is
lower than that derived above, which may be due to the variation of
}$\Theta_{D}$\textcolor{black}{ with temperature.}\textcolor{blue}{
}\textcolor{black}{\ Around $\thickapprox$ 30 K the fit for Y$_{2}$CoGa$_{8}$
is relatively poor for which a possible reason is mentioned below. }

Fig. 1(b) shows the resistivity curves for the Y$_{2}$CoGa$_{8}$
compound with current (J) parallel to the crystallographic directions
{[}100{]} and {[}001{]},\textcolor{blue}{ }\textcolor{black}{\ respectively.
A}long both the directions the temperature dependent resistivity demonstrates
\textcolor{black}{a} metallic behavior,\textcolor{black}{ \ the resistivity
decreases linearly at high }temperatures followed by \textcolor{black}{a
nearly temperature independent behavior below 15 K. The observed }behavior
is \textcolor{black}{in tune with the phonon-induced scattering of
the charge carriers}\textcolor{blue}{ }\ expected in a non-magnetic
compound. \textcolor{black}{There }occurs a hump between 20 and 35
K \textcolor{black}{which is more prominent }along {[}001{]} direction.
\textcolor{black}{A similar hump was also found in polycrystalline
Y$_{2}$CoIn$_{8}$\citet{Devang1}. Such a hump in the resistivity
of a nonmagnetic compound is rarely seen and may arise due to a charge
density wave induced formation of an anisotropic energy gap in the
Fermi surface. Similar behavior is also seen in case of 2H-NbSe$_{2}$,
Nb$_{3}$Te$_{4}$, 2H-TaSe$_{2}$, NbSe$_{3}$, ZrTe$_{3}$, LaAgSb$_{2}$,
etc\citet{Sekine,Craven,Soto,Felser,Myers}. The absence of hysteresis
in resistivity indicates a second order nature of the proposed charge
density wave transition. The prominent effect along the {[}001{]}
may be due to a larger gap along this direction. The deviation of
the fit based on the Debye formula to the heat capacity around 30
K (Fig. 1(a)) may also be due to the same effect.} The resistivity
along {[}001{]} direction \textcolor{black}{is} found to be lower
compared to the in-plane {[}100{]} resistivity. \textcolor{black}{The
similar anisotropic behavior in the resistivity was found for all
the compounds described below and may arise due to the inherent structural
anisotropy of the compound.} The residual resistivity along {[}100{]}
and {[}001{]} directions is 42 $\mu\Omega\,\mathrm{cm}$ and 54 $\mu\Omega\,\mathrm{cm}$
respectively. \textcolor{black}{Overall, th}e resistivity values are
higher compared to corresponding \textcolor{black}{polycrystalline}\textcolor{blue}{
}\ indide Y$_{2}$CoIn$_{8}$\citet{Devang1}. It may be due to the
low density of states at \textcolor{black}{the Fe}rmi level in Y$_{2}$CoGa$_{8}$.
A similar observation on Lu$_{2}$CoGa$_{8}$ would have strengthened
our conjucture but single crystals of Lu$_{2}$CoGa$_{8}$ were too
small for resistivity measurements.

\subsection{Gd$_{2}$CoGa$_{8}$}

The results on Gd$_{2}$CoGa$_{8}$ are presented next as Gd is an
\textit{S}-state ion and the CEF effects are negligible in the first
order approximation. Fig. 2(a) shows the temperature dependence of
resistivity for Gd$_{2}$CoGa$_{8}$ with current parallel to {[}100{]}
and {[}001{]}\textcolor{blue}{ }\textcolor{black}{directions, respectively}\textcolor{blue}{.
}\textcolor{black}{The inset shows the expanded low temperature pa}rt
below 37 K. The resistivity shows a metallic behavior with temperature
down to 35 K.\textcolor{black}{ \ Similar to Y$_{2}$CoGa$_{8}$
t}he resistivity with current along {[}100{]} direction \textcolor{black}{is}
higher than along {[}001{]}. Below 35 K the resistivity with current
parallel to {[}100{]} levels off followed\textcolor{black}{ \ by
a minor kink at T$_{N}$ = 20 K and then it drop}s almost linearly
\textcolor{black}{down to 2 K}. \textcolor{black}{The upward kink
at T$_{N}$ can be attributed to a small gap introduced in the Fermi
surface due to the magnetic super-zone effect\citet{Elliot}.} The
resistivity with current along {[}001{]} direction \textcolor{black}{shows
an apparently anomalous behavior. It increases below 35 K in the paramagnetic
region well above the ordering temperature followed by a sharp increase
at T$_{N}$ and then decreases at low temperatures. The decrease is
not as sharp as expected from the loss in spin disorder resistivity
but tends to fall slowly. The sharp rise at T$_{N}$ is due to the
dominant super-zone gap effect along {[}001{]}. The rise in resistivity
in the paramagnetic state as the temperature approaches T$_{N}$,
is most likely due to the short range antiferromagnetic correlations
\citet{Balberg,Suezaki}. It has been shown theoretically that the
temperature derivative of resistivity has a negative divergence as
T$_{N}$ is approached from the paramagnetic regime due to the large
angle scattering and the divergence in the spin-spin correlation function
at N$\grave{\mathrm{e}}$el temperature \citet{Suezaki} and in reality
it is the life time of the anomalous fluctuation is masked by the
energy transfer from the conduction electrons resulting in the increase
in the resistivity of the compound. Some antiferromagnetic compounds
are known to show similar behavior; for example: Tb \citet{Meaden},
(RPd$_{3}$)$_{8}$Al (R = Tb and Gd) \citet{Surjeet Singh}, UNiAl
\citet{Sechovsky}.} \textcolor{black}{When a magnetic field of 30
kOe is applied along the {[}100{]} direction, the upturn in the resistivity
in the paramagnetic state between $\thickapprox30$ K and T$_{N}$
is partially suppressed as shown in the inset of Fig. 2(a). The suppression
of the upturn in resistivity with field supports the short range antiferromagnetic
correlation in the compound along the {[}001{]} axis. Above 30 K,
the crossover in the H = 0 and 30 kOe plots is likely due to the positive
cyclotron contribution of the charge carriers to magneto-resistance. }

\begin{figure}
\includegraphics[width=0.4\textwidth]{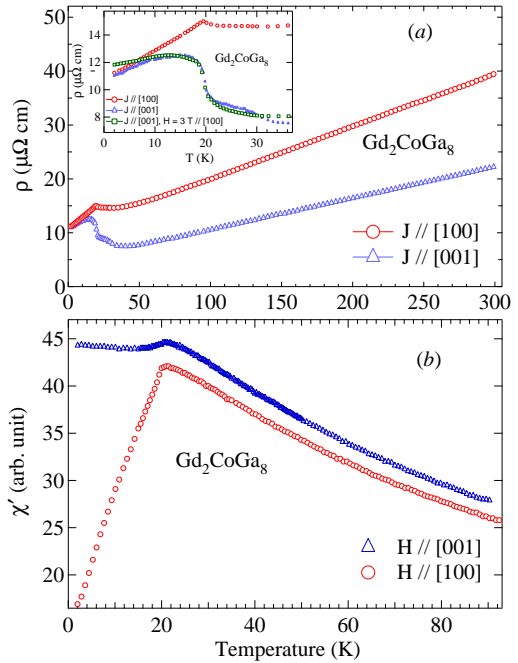}\caption{(a) Resistivity of Gd$_{2}$CoGa$_{8}$ with current parallel to {[}100{]}
and {[}001{]} direction. The inset shows the expanded low temperature
part. (b) \textit{AC} susceptibility for the same with \textit{AC}
field parallel to {[}100{]} and {[}001{]} directions.}

\end{figure}

\textcolor{black}{The anisotropic behavior of resistivity in the neighborhood
of T$_{N}$ in contrast to the corresponding isotropic behavior of
}\textit{DC}\textcolor{black}{ magnetization (5 kOe)\citet{Devang}
motivated us to investigate Gd$_{2}$CoGa$_{8}$ \ with AC susceptibility.
The data are shown in Fig.2(b), with the }\textit{AC}\textcolor{black}{
field applied along the two crystallographic directions. \textgreek{q}\textasciiacute{}
along {[}100{]} increases in the paramagnetic state followed by a
peak at T$_{N}$ and then decreases as expected for a simple antiferromagnet.
On the other hand, along {[}001{]} direction \textgreek{q}\textasciiacute{}
does not decrease below T$_{N}$ and shows a slight upturn at low
temperatures, indicating the presence of complicated magnetic structure
with anisotropy. The interaction of the charge carriers with anisotropic
magnetic configuration is responsible for the observed behavior of
resistivity around T$_{N}$. }

\textcolor{black}{The transverse magnetoresistance (MR), defined as
MR = {[}R(H)-R(0){]}/R(0), of the compound at 2 K with current applied
along the two principal crystallographic directions shows significant
anisotropy (Fig. 3(a)). With current along {[}100{]} the MR increases
almost linearly with field up to approximately 12 \% at 90 kOe, where
as along {[}001{]} direction it varies more strongly increasing nonlinearly
up to 57 \% at 90 kOe. The contribution to the total MR due to spin-orbit
coupling will be negligible for Gd$^{3+}$ ions. The cyclotron contribution
will also not give rise to such a large MR. Field induced metamagnetic
transitions can give rise to a large MR, but Gd$_{2}$CoGa$_{8}$
does not show any metamagnetic behavior at 2 K and further the magnetic
isotherms at 2 K along both the directions nearly coincide with each
other \citet{Devang} thereby pointing out that the magnetoresistivity
behavior in Gd$_{2}$CoGa$_{8}$ \ is primarily influenced by other
factors. Further with both current and field constrained to ab plane
(J // {[}100{]} and H // {[}010{]}) but transverse to each other,
the magnetoresistance increases to 5 \% at 90 kOe. Hence the direction
of the field does not play a major role for anomalously high magnetoresistance
with current parallel to {[}001{]} and only the direction of the current
matters. We suggest that the anisotropy in MR arises due to the anisotropy
of the Fermi surface.}\textcolor{blue}{ }%
\begin{figure}
\includegraphics[width=0.4\textwidth]{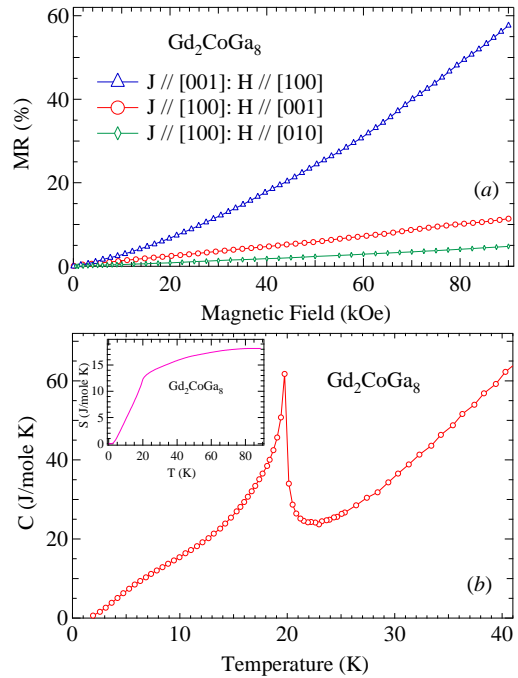}\caption{(a)Mangetoresistance of Gd$_{2}$CoGa$_{8}$ with current parallel
to {[}100{]} and {[}001{]} directions. (b) Heat capacity of Gd$_{2}$CoGa$_{8}$
with inset showing the calculated magnetic entropy.}

\end{figure}
 The heat capacity behavior of Gd$_{2}$CoGa$_{8}$ is shown in Fig.
3(b). It undergoes a lambda type second order magnetic transition
at T$_{N}$ = 20 K consistent with the magnetic susceptibility and
resistivity \textcolor{black}{data.} The magnetic contribution to
the heat capacity was isolated using the data for Lu$_{2}$CoGa$_{8}$,
assuming it as a measure of phonon contribution, taking into account
the mass difference between Gd and Lu. The magnetic entropy calculated
as a function of temperature is shown in the inset of Fig. 3(b).\textcolor{blue}{
}\textcolor{black}{The entropy at T$_{N}$ is 13 J/mole K and it attains
the theoretical value of Rln8 (17.3 J/mole K) at about 60 K. It nearly
saturates above 60 K. This indicates the presence of short range antiferromagnetic
correlations above T$_{N}$ and provides further support to our explanation
of the upturn in the resistivity in the paramagnetic state as mentioned
above. }

\subsection{Tb$_{2}$CoGa$_{8}$, Dy$_{2}$CoGa$_{8}$ and Ho$_{2}$CoGa$_{8}$}

\textcolor{black}{We now describe our results for compounds in which
CEF effects are operative. In ref.1\citet{Devang}, it was found that
for compounds with Tb, Dy and Ho, the easy axis of magnetization is
along {[}001{]}. The resistivity of Tb$_{2}$CoGa$_{8}$ with current
parallel to {[}100{]} and {[}001{]} directions, respectively, is shown
in Fig. 4(a). The resistivity along both the directions initially
decreases linearly with temperature down to $\approx$ 130 K followed
by a relatively faster drop at lower temperatures, which we attribute
to the CEF effect. The thermally induced variation of the fractional
Boltzmann occupation of the CEF levels changes the otherwise constant
spin disorder resistivity. Overall, the decreases in the resistivity
between 1.8 and 300 K is more prominent for J // {[}100{]} ($\approx200$
$\mu\Omega$ cm) than for J // {[}001{]} ($\approx30$ $\mu\Omega$
cm), indicating a significant anisotropy in the transport property
of the compound. A change in the slope at T$_{\mathrm{N}}$ for J
// {[}001{]} (see, inset) occurs due to the loss of spin disorder
resistivity. There is no discernible anomaly at T$_{N}$ for J //
{[}100{]} but a change in slope exists below 15 K. To investigate
a possible origin of this feature, the AC susceptibility was measured
with AC field along {[}100{]} and {[}001{]} directions as shown in
Fig. 4(b). It decreases monotonically below T$_{N}$ along {[}001{]}
but along the {[}100{]} direction it increases below 15 K followed
by a peak at $\approx$ 6 K. The change in the slope of resistivity
at 15 K thus appear to be correlated to the behavior of the AC susceptibility
below 15 K. It is possible that because of some complicated magnetic
structure there is a component along the }\textit{\textcolor{black}{ab}}\textcolor{black}{\ -plane
whose }variation with temperature affects the variation of resistivity.%
\begin{figure}
\includegraphics[width=0.4\textwidth]{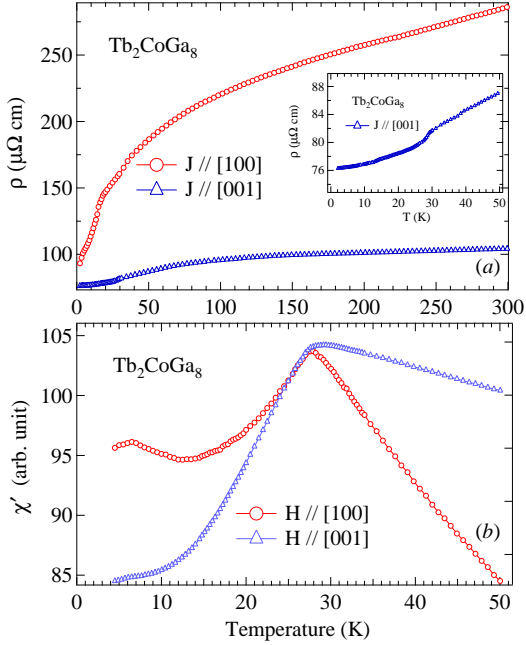}

\caption{(a) Resistivity of Tb$_{2}$CoGa$_{8}$ with current parallel to {[}100{]}
and {[}001{]} direction\textcolor{black}{s, respectively. }(a) \textit{\textcolor{black}{AC}}
\ susceptibility of Tb$_{2}$CoGa$_{8}$ with AC field parallel to
{[}100{]} and {[}001{]} directions, respectively.}

\end{figure}
\textcolor{black}{The magnetoresistance of the compound with current
parallel to {[}100{]} and {[}001{]} directions, respectively, is shown
in Fig. 5(a). The positive magnetoresistance along both the crystallographic
directions is consistent with the antiferromagnetic behavior of the
compound. Magnetoresistance at 2 K with H // {[}100{]} and J // {[}001{]}
increases almost linearly by 17 \% with applied field increased to
90 kOe. On the other hand the magnetoresistance shows a complex behavior
for H // {[}001{]}. At 2 K for J // {[}100{]}, MR initially increases
linearly with field. There is a rapid increase in a narrow interval
near H $\sim$ 35 kOe followed by a distinct change in the variation
near 82 kOe. The magnetoresistance at 90 kOe is $\approx$ 77 \%,
which though appreciable is far less than that of the corresponding
indide Tb$_{2}$CoIn$_{8}$($\sim$ 2700 \% at 2 K). The anomalies
at 42 and 82 kOe are consistent with the metamagnetic transitions
observed in the magnetic isotherm of the compound at 2 K along the
easy axis as reported in ref.1\citet{Devang}. At 5 K, the magnetoresistance
decreases but qualitatively the behavior is similar to that at 2 K.
Increasing the temperature to 10 K, the first anomaly in MR is slightly
shifted up in field whereas above 80 kOe the magnetoresistance decreases
with field. The latter is due to the reduction in the scattering of
the conduction electrons by the ferromagnetically aligned Tb$^{3+}$
ions at high fields where the compound enters the field induced ferromagnetic
state. Here the field-induced polarized state is achieved by the combined
action of field and temperature ($\sim$ 80 kOe and 10 K, respectively).
}%
\begin{figure}
\textcolor{black}{\includegraphics[width=0.4\textwidth]{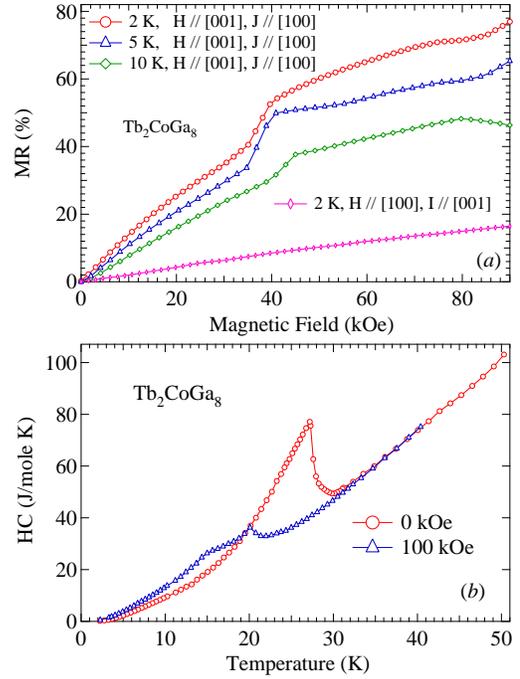}\caption{(a) Magnetoresistance for Tb$_{2}$CoGa$_{8}$ with current and field
in the indicated direction. (b) Heat Capacity for the same with and
without field.}
}

\end{figure}
\textcolor{black}{Higher fields are required to induce the ferromagnetic
state at lower temperatures. The heat capacity of the Tb$_{2}$CoGa$_{8}$
(Fig. 5(b)) in zero field is dominated by a lambda type anomaly at
the N$\acute{\mathrm{e}}$el temperature (T$_{N}$= 27.5 K). In an
applied field of 100 kOe the anomaly disappears; a broad hump and
a kink appear at lower temperatures, reflecting an overall weakning
of the antiferromagnetic configuration in applied fields and field
induced metamagnetic transition in the compound.}

\textcolor{black}{}%
\begin{figure}
\includegraphics[bb=0bp 0bp 639bp 412bp,width=0.4\textwidth]{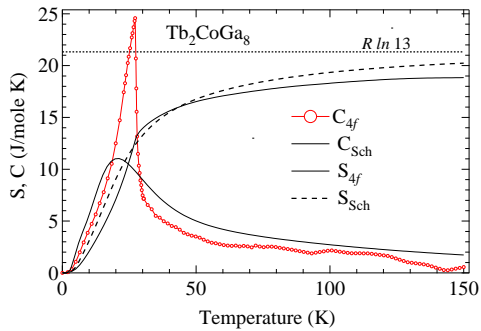}

\textcolor{black}{\caption{\textcolor{black}{The 4}\textit{\textcolor{black}{f}} contribution
\textcolor{black}{C$_{4f}$} to the heat capacity of Tb$_{2}$CoGa$_{8}$with
Schottky contribution \textcolor{black}{C$_{Sch}$} estimated from
CEF split energy levels. The corresponding entropies estimated from
\textcolor{black}{4}\textit{\textcolor{black}{f}} contribution and
CEF split levels are also shown.}
}

\end{figure}
\textcolor{black}{ The 4}\textit{\textcolor{black}{f}}\textcolor{black}{
\ contribution to the heat capacity, C$_{4f}$, determined using
the same procedure as mentioned for Gd$_{2}$CoGa$_{8}$, and the
entropy S$_{4f}$ is shown in Fig. (6). In addition we have also plotted
the Schottky specific heat C$_{Sch}$ and the corresponding entropy
S$_{Sch}$ calculated from the following expressions }

\textcolor{black}{\begin{equation}
C_{Sch}=\frac{\partial}{\partial T}\left[\frac{1}{Z}\sum_{n}E_{n}exp\left(-\frac{E_{n}}{k_{B}T}\right)\right]\end{equation}
}

\textcolor{black}{\begin{equation}
S_{4f,\, Sch}=\intop_{0}^{T}\,\frac{C_{4f,\, Sch}}{T}dT\end{equation}
where, }\textit{\textcolor{black}{Z }}\textcolor{black}{is the partition
function, }\textit{\textcolor{black}{E}}\textcolor{black}{$_{n}$
are the CEF split energy levels derived from the CEF fitting of the
inverse magnetic susceptibility in ref.1\citet{Devang}. It is evident
from the figure that there is a reasonably good agreement between
C$_{4f}$ and C$_{Sch}$ in the paramagnetic regime. This s}upports
the validity of the CEF level scheme for T\textcolor{black}{b$_{2}$CoGa$_{8}$
as derived from the magnetization data. \ The entropy obtained from
the magnetic contribution to the heat capacity is 18.9 J/mole K at
150 K. The theoretically expected value of $R\, ln\left(2J+1\right)$(21.32
J/mole K) will be achieved at higher temperatures when all the CEF
levels are thermally populated. }

\begin{figure}
\includegraphics[width=0.4\textwidth]{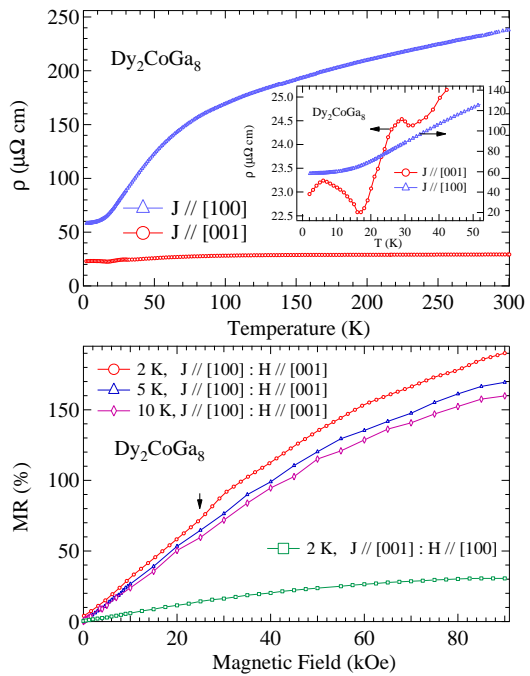}\caption{(a)\textcolor{black}{ Resistivity of Dy$_{2}$CoGa$_{8}$ with current
along the two crystallographic directions, respectively; the inset
shows a magnified view below 50 K; }(b)\textcolor{black}{ Magnetoresistance
at selected temperatures for configuration as mentioned in the figure.}}

\end{figure}
\textcolor{black}{The temperature dependence of electrical resistivity
from 1.8 to 300 K for Dy$_{2}$CoGa$_{8}$ is shown in Fig. 7(a).
The high temperature part of the electrical resistivity of Dy$_{2}$CoGa$_{8}$
is qualitatively similar to that of Tb$_{2}$CoGa$_{8}$ along the
two crystallographic directions, whereas the low temperature resistivity
shows a different behavior. The resistivity for J // {[}100{]} decreases
monotonically below 50 K and does not show any anomalies and becomes
nearly temperature independent below 20 K. Along {[}001{]} the resistivity
shows peaks at $\approx$ 29 and 6 K, the former in the paramagnetic
state and the latter below T$_{N}$. The increase in the resistivity
below T$_{N}$ (17 K) is due to the super-zone gap effects, and it
is also consistent with the magnetization results which show that
the moments order along the {[}001{]} direction. In order to look
for possible origin of the paramagnetic peak at 29 K, we measured
the AC susceptibility with field along {[}001{]} direction (not shown).
However, no anomaly was found in AC susceptibility; neither do we
observe any anomaly at 29 K in the heat capacity (may be overridden
by the magnetic contribution) (Fig. 8(a)).} %
\begin{figure}[b]
\includegraphics[width=0.4\textwidth]{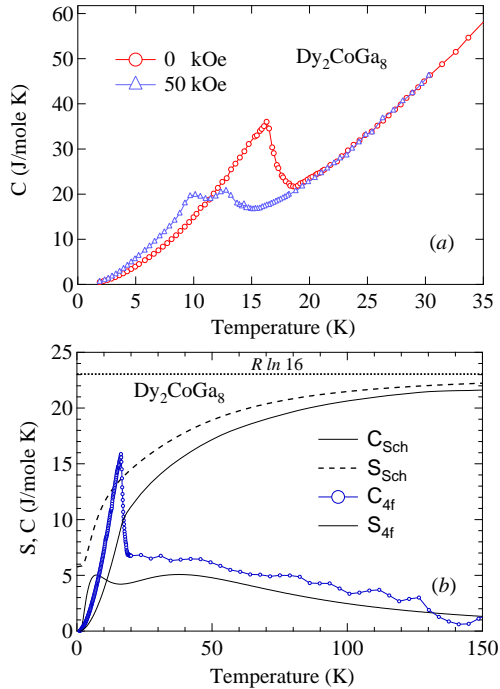}\caption{(a) Heat capacity of Dy$_{2}$CoGa$_{8}$. (b) \textcolor{black}{The
4}\textit{\textcolor{black}{f}} \ contribution to the heat capacity
of Tb$_{2}$CoGa$_{8}$with Schottky fit estimated from CEF split
energy levels. The entropy estimated from magnetic contribution and
CEF split levels are also shown.}

\end{figure}
 \textcolor{black}{The sharp drop in resistivity below $\approx$
29 K (Fig. 7(a) inset) with decreasing temperature rules out the possibility
of spin fluctuation. The $\approx$ 29 K peak in Dy$_{2}$CoGa$_{8}$
exists nearly at the same temperature as in Y$_{2}$CoGa$_{8}$. Hence
it may also be due to a charge density wave induced gap in the Fermi
surface as speculated for Y$_{2}$CoGa$_{8}$. However, this needs
further investigation. The magnetoresistance at 2 K with J // {[}001{]}
and H // {[}100{]} as depicted in Fig. 7(b) is similar to Tb$_{2}$CoGa$_{8}$
and it increases to 25 \% at 90 kOe. For J // {[}100{]} and H // {[}001{]}
there is a change in slope at 24 kOe (shown by an arrow) above which
the magnetoresistance increases monotonically to 190 \% at 90 kOe.
With increase in the temperature the magnetoresistance decreases and
shows a negative curvature indicating the onset of field induced ferromagnetic
behavior. The heat capacity of Dy$_{2}$CoGa$_{8}$ is shown in Fig
8(a). A lambda type anomaly indicates the magnetic transition. Application
of a magnetic field of 50 kOe results in two humps. The effect is
similar to that observed in Tb$_{2}$CoGa$_{8}$. The 4}\textit{\textcolor{black}{f}}\textcolor{black}{
\ contribution to the heat capacity of Dy$_{2}$CoGa$_{8}$ and the
Schottky curve calculated as explained above are shown in Fig. 8(b).
In the paramagnetic regime, C$_{Sch}$ and C$_{4f}$ are in fair qualitative
agreement with each other. S$_{4f}$ and S$_{Sch}$ are seen to approach
the theoretically expected value at high temperatures. It may be mentioned
here that Dy is a Kramer's ion, and the CEF levels in the tetragonal
point symmetry will split into 8 doublets\citet{Devang}. But the
calculated C$_{Sch}$ and S$_{Sch}$ do not take into account the
contribution (}\textit{\textcolor{black}{Rln}}\textcolor{black}{2)
from the doublet ground state. Therefore, we have shifted up our plot
of S$_{Sch}$ up by $Rln2$. }

Ho$_{2}$CoGa$_{8}$ orders antiferromagnetically at 6 K with easy
axis of magnetization along the {[}001{]} direction\citet{Devang}.
\textcolor{black}{The resistivity of the compound is shown in Fig.
9(a) with the magnified low temperature part as an inset. The resistivity
shows a drop at the ordering temperature of the compound with current
parallel to {[}100{]} and {[}001{]} directions.} %
\begin{figure}
\includegraphics[width=0.4\textwidth]{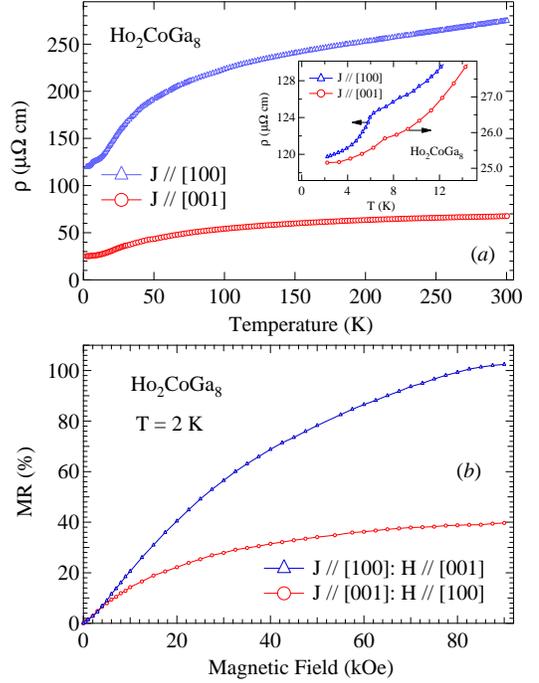}

\caption{(a) Resistivity of Ho$_{2}$CoGa$_{8}$ with current parallel to \textcolor{black}{{[}100{]}
and {[}001{]}, respectively with inset showing the magnified low temperature
part}. (b) Magnetoresistance at 2 K with current and field parallel
to the indicated directions.}

\end{figure}
\textcolor{black}{The overall electrical resistivity with J // {[}100{]}
is similar to that observed by Adriano et al.\citet{Adriano}. Similar
to the other members of the series, the resistivity with current parallel
to {[}100{]} is higher than with the current parallel to {[}001{]}.
Below 100 K CEF effects manifest in a relatively faster decrease of
the resistivity with temperature along both the directions. The magnetoresistance
of the compound is shown in Fig. 9(b) with the indicated direction
of current and field. The magnetoresistance with J // {[}100{]} and
H // {[}001{]} increases with field up to $\approx$ 102 \% at 90
kOe while the corresponding variation with J // {[}001{]} and H //
{[}100{]} is $\approx$ 40 \%. The high magnetoresistance with H //
{[}001{]} is in agreement with the easy axis of magnetization {[}001{]}
of the compound. The heat capacity of Ho$_{2}$CoGa$_{8}$ plotted
in Fig. 10(a) shows a sharp lambda type anomaly at the magnetic ordering
temperature of the compound. The peak shifts to lower temperatures
in an applied field of 30 kOe as anticipated for an antiferromagnetically
ordered compound. On further increasing the field to 100 kOe the peak
disappears completely and the heat capacity shows a large hump centered
at 8 K. Both these effects most likely arise from the metamagnetic
transition in the compound. The 4}\textit{\textcolor{black}{f}}\textcolor{black}{
\ contribution to the heat capacity of Ho$_{2}$CoGa$_{8}$ and the
Schottky curve is shown in Fig. 10(b). The Schottky curve shows a
peak and hump in fair agreement with the experimental curve. The estimated
entropies from the 4}\textit{\textcolor{black}{f}}\textcolor{black}{
\ contribution to the heat capacity and the Schottky energy levels
are almost equal to the theoretically expected value of $Rln17$.}

\begin{figure}[h]
\includegraphics[width=0.4\textwidth]{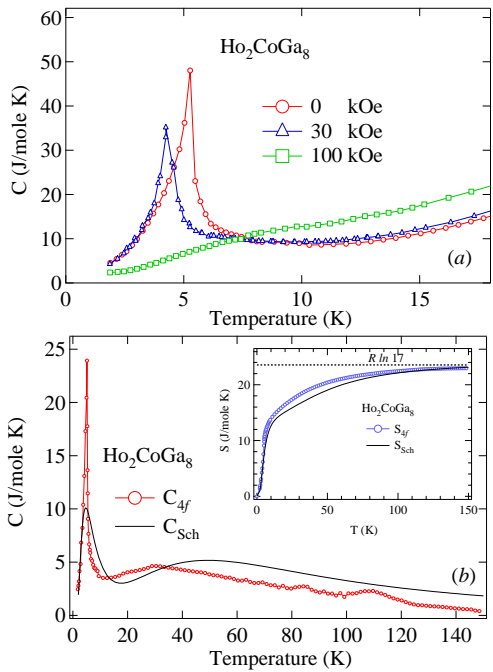}

\caption{(a)\textcolor{black}{ Heat Capacity of Ho$_{2}$CoGa$_{8}$ in applied
fields of 0, 30 and 100 kOe. }(a)\textcolor{black}{ The 4}\textit{\textcolor{black}{f}}\textcolor{black}{
\ contribution to the heat capacity of Ho$_{2}$CoGa$_{8}$ and the
Schottky heat capacity curve estimated from CEF split energy levels.
The entropies estimated from the 4}\textit{\textcolor{black}{f}}\textcolor{black}{
\ contribution and CEF split levels are shown in the inset.}}

\end{figure}

\subsection{Er$_{2}$CoGa$_{8}$ and Tm$_{2}$CoGa$_{8}$}

In case of Er$_{2}$CoGa$_{8}$ and Tm$_{2}$CoGa$_{8}$ the easy
axis of magnetization is along the \textit{ab}\ -plane, unlike the
other compounds described above where the easy axis of magnetization
was along the {[}001{]} direction. These \textcolor{black}{two}\textcolor{blue}{
}\ compounds order antiferromagnetically at 3 K and 2 K, respectively\citet{Devang}.
The resistivity of both the compounds is shown in Fig. 11. For Er$_{2}$CoGa$_{8}$
(Fig. 11(a)) the resistivity decreases linearly from room temperature
as expected for a metallic compound down to 100 K. Below 100 K\textcolor{blue}{
}\textcolor{black}{\ the faster drop is attributed }to crystal field
effects. The inset shows the low temperature part of the resistivity.
When J // {[}100{]} the resistivity increases below $\approx$ 6 K
followed by the downward drop for T < T$_{N}$ and with \textcolor{black}{J
// }{[}001{]} the resistivity falls below $\approx$ 6 K. %
\begin{figure}[h]
\includegraphics[width=0.4\textwidth]{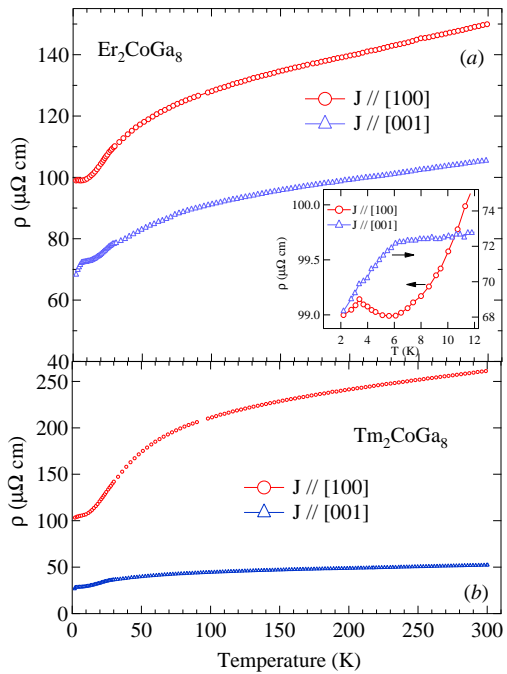}

\caption{(a)\textcolor{black}{ Resistivity of Er$_{2}$CoGa$_{8}$ with current
parallel to {[}100{]} and {[}001{]} directions respectively; the inset
shows the low temperature part, }(b)\textcolor{black}{ Resistivity
of Tm$_{2}$CoGa$_{8}$ with current parallel to {[}100{]} and {[}001{]}
directions. }}

\end{figure}
\textcolor{black}{The upturn in the resistivity with J // {[}100{]}
below 6 K can occur due to the presence of short range antiferromagnetic
correlations along the }\textit{\textcolor{black}{ab}}\textcolor{black}{
plane which incidentally is the easy axis of magnetization. The drop
in the resistivity at T$_{N}$ (3 K) is due to disappearance of spin
disorder resistivity. The electrical resistivity for J // {[}001{]}
direction shows a drop exactly at the same temperature (6 K) where
there was an increase in the resistivity along the other direction.
Although the exact reason for this behavior is not known at present,
we tentatively attribute it to the short range correlations of the
moments along the easy axis. These relatively opposite variations
in the thermal variation of the resistivity indicate that the configurations
of the moments when resolved along different directions can be different.
Compared to relatively simple ferromagnets, magnetic moments in antiferromagnetic
compounds can have very complicated alignments described by a number
of wave vectors, phase angles, etc. In case of Tm$_{2}$CoGa$_{8}$
the high temperature resistivity has the behavior similar to that
of Er$_{2}$CoGa$_{8}$ and at low temperatures the resistivity falls
at the Neel temperature (2 K) of the compound. The magnetoresistance
at 2 K for Er$_{2}$CoGa$_{8}$ and Tm$_{2}$CoGa$_{8}$ is shown
in Fig. 12. }%
\begin{figure}[h]
\includegraphics[width=0.4\textwidth]{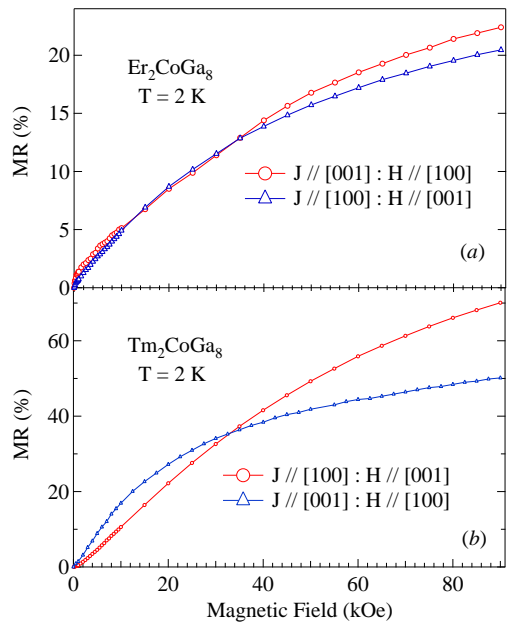}

\caption{Magnetoresistance of Er$_{2}$CoGa$_{8}$ and Tm$_{2}$CoGa$_{8}$
at 2 K with current and field parallel to the indicated directions.}

\end{figure}
 \textcolor{black}{The magnetoresistance for Er$_{2}$CoGa$_{8}$
with H // {[}100{]} and {[}001{]} respectively are close to each other
up to 30 kOe. At higher fields the magnetoresistance with H // {[}100{]}
is marginally higher and attains a maximum value of $\approx$ 23
\% at 90 kOe. The result is consistent with lesser anisotropic magnetic
behavior seen in the magnetization data; also the MR is measured close
to the ordering temperature of the compound. In case of Tm$_{2}$CoGa$_{8}$
the magnetoresistance with J // {[}001{]} and H // {[}100{]} is higher
than that with J // {[}100{]} and H // {[}001{]} below 30 kOe but
at higher fields the latter dominates up to the highest applied field.}
\begin{figure}[h]
\includegraphics[width=0.4\textwidth]{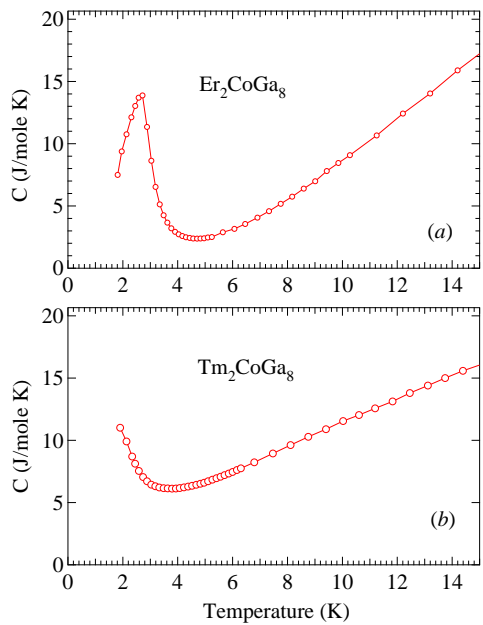}

\caption{Heat capacity of Er$_{2}$CoGa$_{8}$ and Tm$_{2}$CoGa$_{8}$ }

\end{figure}
The magnetoresistivity at 90 kOe with H // {[}001{]} is 70 \% and
50 \% for H // {[}100{]}. The initial high value of magnetoresistance
for J // {[}001{]} and H // {[}100{]} can be understood by the fact
that {[}100{]} is the easy axis of magnetization for the compound.
Above $\approx$ 30 kOe the magnetoresistance\textcolor{black}{ \ tends}
to saturate with field along {[}100{]} direction indicating the development
of the ferromagnetic component (-ve component). \textcolor{black}{With
field along the hard direction {[}001{]} the development of ferromagnetic
component will be achieved at higher fields and because of this the
magnetoresistance increases leading to the observed crossover. The
heat capacity of both the compounds is shown in Fig. 13. Er$_{2}$CoGa$_{8}$
shows an anomaly at the ordering temperature (3 K) while the up turn
below 3 K is precursor to the magnetic transition at 2 K in Tm$_{2}$CoGa$_{8}$.
Since the heat capacity is still large at the lowest temperature data
point in Er$_{2}$CoGa$_{8}$ and it is necessary to have the heat
capacity data to lower temperatures in both the Er and Tm compounds,
it is not possible to calculate S$_{4f}$ in these two compounds.
The 4}\textit{\textcolor{black}{f}}\textcolor{black}{ \ contribution
to the heat capacity of Tm$_{2}$CoGa$_{8}$ and the Schottky curve
is shown in Fig. 14. The Schottky curve shows a low temperature rise
and hump in fair agreement with the experimental curve. A similar
analysis for Er$_{2}$CoGa$_{8}$ is not shown because the C$_{Sch}$
calculated from the CEF energy levels as deduced from the magnetization
data {[}1{]} did not match well with 4}\textit{\textcolor{black}{f}}\textcolor{black}{
\ contribution to the heat capacity. }%
\begin{figure}[H]
\includegraphics[width=0.4\textwidth]{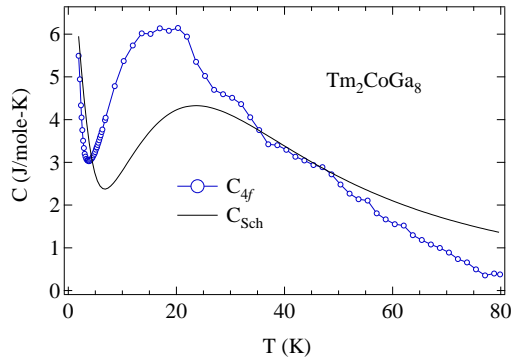}\caption{\textcolor{black}{The 4}\textit{\textcolor{black}{f}}\textcolor{black}{
\ contribution to the heat capacity and the Schottky heat capacity
of Tm$_{2}$CoGa$_{8}$.}}

\end{figure}

\subsection{Conclusion}

To conclude, we have studied the anisotropic electrical resistivity,
magnetoresistance and heat capacity of R$_{2}$CoGa$_{8}$ single
crystals.\textcolor{black}{ \ Heat capacity data provide evidence
of bulk magnetic transitions with ordering temperatures matching with
our earlier magnetization studies. The effect of external magnetic
field on the heat capacity of these compounds is in conformity with
their antiferromagnetic nature. The Schottky heat capacity calculated
from the CEF energy levels derived from the magnetization data compares
well with 4}\textit{\textcolor{black}{f}}\textcolor{black}{ \ contribution
to the heat capacity in the paramagnetic regime for R = Tb, Dy, Ho
and to a lesser extent in Tm$_{2}$CoGa$_{8}$, thus strengthening
the validity of the CEF level scheme obtained in ref. {[}1{]} for
these compounds. The electrical resistivity shows a significant anisotropy,
the resistivity along the }\textit{\textcolor{black}{ab\ }}\textcolor{black}{
plane being higher compared to its magnitude along the }\textit{\textcolor{black}{c}}\textcolor{black}{\ -axis.
This anisotropic transport behavior indicates a dominant electron
motion along the }\textit{\textcolor{black}{c}}\textcolor{black}{\ -axis
and may arise due to the structural anisotropy of the compound. Anomalous
upturn in the electrical resistivity of some compounds as T$_{N}$
is approached from the paramagnetic side is attributed to short range
antiferromagnetic correlations. A hump in the non-magnetic Y$_{2}$CoGa$_{8}$
and Dy$_{2}$CoGa$_{8}$ near 29 K is tentatively attributed to CDW
ordering, which needs to be probed further for confirmation. The highly
anisotropic magnetoresistance data reflect the effect of metamagnetic
transition in these compounds. }

\end{document}